\documentclass[11pt,twocolumn]{article}
\usepackage{abstract,footmisc}
\usepackage{wide,graphics,wrapfig}
\usepackage{filecontents}
\usepackage{pstricks}
\usepackage{pst-bar}
\usepackage{pstricks-add}
\usepackage{pst-all}
\newcommand{\remove}[1]{}

\definecolor{mygrey0}{rgb}{0.91,0.91,1.00}
\usepackage{pifont}\newcommand{\qed}{\hfill\ding{113}}

\def\bfp#1{(\textbf{#1})}
\def\itp#1{(\textit{#1}\/)}
\newtheorem{lemma}{Lemma}

\newenvironment{proof}{\begin{trivlist}%
 \item[\hspace{\labelsep}{\bf\noindent Proof: }]}{\hfill\qed\end{trivlist}}
\makeatletter
\SpecialCoor
\def\errorLine{\@ifnextchar[{\pst@errorLine}{\pst@errorLine[]}}
\def\pst@errorLine[#1](#2)#3#4{{%
    \ifx#1\empty\else\psset{#1}\fi
    \pst@getcoor{#2}\pst@tempb
    \def\@errorMin{#3}
    \def\@errorMax{#4}
    \psline{|-|}%
    (!%
        /yDot \pst@tempb exch pop \pst@number\psyunit div def
        /xDot \pst@tempb pop \pst@number\psxunit div def
        xDot yDot \@errorMin\space add%
    )(!%
        /yDot \pst@tempb exch pop \pst@number\psyunit div def
        /xDot \pst@tempb pop \pst@number\psxunit div def
        xDot yDot \@errorMax\space add%
    )
}}
\def\GetCoordinates#1{\expandafter\GetCoordinates@i#1}
\def\GetCoordinates@i #1{\GetCoordinates@ii#1}
\def\GetCoordinates@ii#1 #2 #3 #4 #5 #6 #7 #8 {%
    \DoCoordinate{#2}{#4}%
    \errorLine[linecolor=black, linewidth=1pt](#2,#4){#6}{#8}
    \@ifnextchar D{\GetCoordinates@ii}{}%
}
\makeatother
  
\begin{document}
\title{Local Read-Write Operations in Sensor Networks\thanks{}}
\author{Ted Herman \\ University of Iowa \\
\texttt{herman@cs.uiowa.edu} \and Morten Mjelde \\ 
University of Bergen \\ 
\texttt{mortenm@ii.uib.no}}
\twocolumn[
\maketitle
\begin{onecolabstract}
Designing protocols and formulating 
convenient programming units of abstraction for
sensor networks is challenging due to communication 
errors and platform constraints.  This paper investigates 
properties and implementation reliability for a 
\emph{local read-write} abstraction.  Local read-write is 
inspired by the class of read-modify-write 
operations defined for shared-memory multiprocessor architectures.  
The class of read-modify-write operations is 
important in solving consensus and related synchronization
problems for concurrency control.  Local read-write is shown
to be an atomic abstraction for synchronizing neighborhood
states in sensor networks.  The paper compares local read-write to 
similar lightweight operations in wireless sensor networks, such 
as read-all, write-all, and a transaction-based abstraction:  for 
some optimistic scenarios, local read-write is a more efficient 
neighborhood operation.  A partial implementation is described,
which shows that three outcomes characterize operation response:
success, failure, and cancel.  A failure response indicates 
possible inconsistency for the operation result, which is the
result of a timeout event at the operation's initiator.  The 
paper presents experimental results on operation performance
with different timeout values and situations of no contention,
with some tests also on various neighborhood sizes.
\end{onecolabstract}
]
\renewcommand{\thefootnote}%
      {\fnsymbol{footnote}}
    \footnotetext[1]{Research supported in part by NSF award 0519907.}
\section{Introduction} \label{section:introduction}
Wireless Sensor Network (WSN) platforms add a twist to traditional 
programming assumptions.  Many resources can be quite constrained,
including bandwidth, program memory, and platform computing 
power.  Not surprisingly, research on sensor network programming
to date has sought abstractions and tools that can satisfy the 
resource constraints, yet enable productivity in software 
development cycles.  Typically, these abstractions are not entirely 
new ideas, but adaptations of (perhaps less orthodox) techniques
from areas of signal processing, database, and parallel or 
distributed computing.  This paper follows the same research direction,
exploring the adaptation of a read-modify-write abstraction as a
unit of sensor network programs;  we propose an operation called
\emph{local read-write} (LRW) for neighborhood communication in 
a sensor network.  
\par
\begin{figure}[h]
\label{fig:simple-lrw}
\begin{minipage}{\columnwidth}
\begin{pspicture}(-0.3,0)(5,3) 
\psframe(0.1,0.1)(5.0,3.0)
\psclip{\psframe[linestyle=none](0.15,0.15)(4.95,2.95)}
  \pscircle[linecolor=gray,fillcolor=mygrey0,fillstyle=solid](0.5,1.5){2.8}
  \endpsclip
\rput(0.8,1.5){\circlenode{x}{\footnotesize $x$}}
\rput(2.5,1.5){\circlenode{y}{\footnotesize $y$}}
\rput(4.5,1.5){\circlenode{z}{\footnotesize $z$}}
\ncarc[arcangle=40]{->}{x}{y}
\ncarc[arcangle=40]{->}{y}{x}
\ncline[linestyle=dashed,linecolor=lightgray]{->}{z}{y}
\rput(1.4,2.8){\tiny \emph{shaded area is}}
\rput(1.4,2.55){\tiny \emph{neighborhood of x}}
\end{pspicture}
\end{minipage}
\caption{\footnotesize LRW with one neighbor.}
\end{figure}
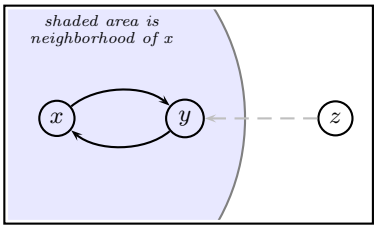

The compare-and-swap (\textsc{c\&s}) instruction, 
available on many multiprocessor
architectures, is an example of read-modify-write.  
In one atomic step, a processor executing \textsc{c\&s} 
conditionally swaps the content of a memory word 
with the content of a register;  the condition for this swap is 
that the content of the memory word have a prescribed value given
as a field of the instruction or given in another register.  
This idea, that a single instruction specifies a condition, a
write value, and expects a response value, can be generalized and
translated to the setting of nodes and packet-based communication.
A simple instance of an LRW operation is illustrated in 
Figure~\ref{fig:simple-lrw}.  Sensor node $x$ initiates the 
operation by transmitting a packet to neighboring node $y$. 
Node $y$ inspects the packet, and possibly 
schedules a tentative write to some local variables; then 
$y$ transmits a response packet to $x$.  Upon receipt of $y$'s
response, node $x$ will either decide to confirm the operation
or back out and void the operation.  Voiding the operation will
result in $y$ discarding its scheduled, tentative write.   
\par
Node $z$ shown in Figure~\ref{fig:simple-lrw} lies outside  
$x$'s neighborhood; there is the possibility that $z$ could
initiate an LRW operation concurrently with $x$, so that $y$ 
first receives a packet from $x$, then a packet from $z$, 
and these requests conflict because they write to the same
location.  For these LRW operations to be atomic, 
the net effect of running both should be logically serial, 
that is, as though one operation completes 
before the other begins.  The design choice for this 
paper is that $y$ should reject $z$'s request while $x$'s operation
is pending, that is, $y$ should immediately send a negative
response to $z$.  
\par
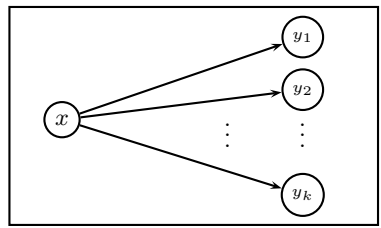
\begin{figure}[h]
\begin{minipage}{\columnwidth}
\begin{pspicture}(-0.3,0)(5,3) 
\psframe(0.1,0.1)(5.0,3.0)
\rput(0.8,1.5){\circlenode{s}{\footnotesize $x$}}
\rput(4.0,2.6){\circlenode{t1}{\tiny $y_1$}}
\rput(4.0,1.9){\circlenode{t2}{\tiny $y_2$}}
\rput(4.0,1.4){\footnotesize $\vdots$}
\rput(3.0,1.4){\footnotesize $\vdots$}
\rput(4.0,0.5){\circlenode{t3}{\tiny $y_k$}}
\ncline{->}{s}{t1}
\ncline{->}{s}{t2}
\ncline{->}{s}{t3}
\end{pspicture}
\end{minipage}
\caption{\footnotesize LRW with $k$ neighbors.}
\label{fig:multi-lrw}
\end{figure}

The single-node neighborhood of $x$, in Figure~\ref{fig:simple-lrw}, 
can be generalized to larger neighborhoods, 
illustrated by Figure~\ref{fig:multi-lrw}.  Node $x$'s LRW operates on
a neighborhood of nodes $y_1$ through $y_k$.  Although the figure
suggests $k$ messages would be transmitted by $x$, a single local
broadcast suffices for many radio platforms.  A typical WSN application 
for LRW is data aggregation.  Suppose each $y_i$ has recorded some 
sensor value $d_i$, and node $x$'s task is to compute some function of 
$\{\,d_i\,|\,1\leq i\leq k\}$ and save the result to its flash memory.
After $x$ has completed this aggregation, each $y_i$ can discard its
$d_i$ value and recycle local memory.  Note that if $y_i$ were to 
asynchronously send $d_i$ to $x$, it could be that $x$ does not have 
local buffers available for this data;  putting $x$ in control is 
a way to manage resources safely.  In one LRW operation, $x$ can 
collect all $d_i$ values and also schedule the $d_i$ variables at
each $y_i$ for recycling.  However, if $x$ does not collect enough
$d_i$ values, say fewer than $k/2$ neighbors respond to the
request initiated within the LRW operation, then $x$ could cancel
the operation and retry it later.  Classical applications of 
read-modify-write, such as consensus or leader election 
(applied to a WSN neighborhood) can easily be expressed as an
LRW operation.
\par

\medskip\par\noindent
\textbf{Contributions and Organization.} 
Section \ref{section:previous-work} summarizes related work. 
Section \ref{section:lrw-design} specifies LRW properties  
and exposes some design choices for implementation.  
Section \ref{section:lrw-theory} 
presents a theoretical result showing how a model based on this 
abstraction differs from other choices.  
Section \ref{section:lrw-experiments} contains 
implementation results, which feature experiments to show 
design tradeoffs.  Discussion of conclusions is in 
Section \ref{section:conclusion}.

\section{Motivation and Related Work} \label{section:previous-work}

Several tracks of WSN research draw analogies to database and 
parallel computing methods.  Early proposals for querying sensor
networks motivated protocols for aggregation and routing to support
query language operations \cite{AY03,GM04,NGSA08}.  
The idea of programming a WSN as a whole
(called macroprogramming) sometimes take the position that programming
sensors resembles the ensemble programming of parallel computing 
machines, using SIMD or MIMD instruction sequences \cite{NMW07,SG08}.  
Inspired by distributed computing research, there are proposals to adapt such
paradigms as snapshots, leader election, and wave computations in 
WSN systems \cite{Y05,HVYL+06,UG07}.  
This paper draws analogy to instructions for 
atomic communication in multiprocessor, shared memory systems.  
\par
In contrast to high-level concepts for WSN software, there is 
also significant research adapting the techniques of \emph{ad hoc}
networks, peer-to-peer, and even internet protocols to the needs 
of WSN applications and the limitations of WSN platforms.  Two
priorities for such research are reliable communication and 
power conservation.  A question emerging from this research is:
what kind of communication abstractions will be convenient for 
programming (\emph{i.e.}, the interfaces are simple and hide low-level
complexity and problems of heterogeneous platforms) while enabling 
efficient use of resource?  This question has predominantly been
investigated with respect to non-local communication, for instance, 
multi-hop protocols, routing structures, and middleware
services for publish-subscribe abstractions.   Our work looks at
local communication, where ``local'' refers to single-hop 
communication, also called \emph{neighborhood} communication.
\par
On one hand, the literature of MAC protocols, specialized to WSN platforms, 
extensively explores the concerns of local communication \cite{A+06}.  
Platform hardware may directly support unicast and neighborhood broadcast
operations, and some radio chips provide low-level support for unicast 
packet acknowledgment in one programmable operation.  On the other hand,    
there are several papers \cite{H03,KA06,D07} suggesting higher-level 
programming units for local communication.  A natural abstraction for 
local communication is \emph{atomic read-all}, which is the operation of
reading the local states of all nodes in a neighborhood.  Using atomic
read-all operations, programs elegantly express calculation of 
neighborhood statistics;  Section \ref{section:lrw-theory} 
elaborates on a variant of read-all with stronger atomicity properties. 
Unfortunately, the read-all abstraction does
not efficiently map to WSN platform abilities.  An alternative abstraction 
is \emph{atomic write-all}, which may be implemented by a single 
local message broadcast.  A write-all operation writes (some part of) 
the states of every other node in a neighborhood.  This operation is not
so natural for programming as read-all, however program transforms
have been proposed that convert many programs using read-all operations
into ones that employ only write-all operations \cite{H03,HT04,KA06}.  
Reliability is a concern with na\"ive implementation of write-all
consisting of a single message broadcast;  
the broadcast can lose messages to a subset of neighbors due to noise 
or collision with other message traffic, say originating from other 
neighborhoods in the WSN (in \cite{KA06}, the basic operation is 
called ``write-all with collision'').   

The concerns of reliability
and atomicity are fundamental to database transaction theory, where
ACID properties define correct transaction processing.  The 
paper \cite{D07} suggests a local WSN operation motivated by 
database transactions:  one atomic operation reads from a subset of
neighbors and writes to a subset of neighbors.  
To improve reliability,   
the local transaction implementation consists
of a sequence of messages: read-request, response, then write-commit 
or abort-transaction.  Transactions may be aborted because
of interference with contending transactions, and the aborted transactions
need to be retried.  The reliability of such local 
transactions is imperfect:  final commit messages can be lost and the 
transaction initiator can crash.  Standard techniques that add reliability
to database transaction processing, such as stable storage and transaction
journaling, are unrealistic for many WSN platforms. 

To give some idea of the resources needed for the operations 
discussed above, Table~\ref{fig:table-ops} summarizes 
\emph{optimistic}, best-case resource measures 
for a neighborhood of $n$ nodes. The two measures are 
number of messages (including both unicast and local broadcast messages)
and number of rounds, where a round is a time interval of sufficient 
length to allow all nodes in a neighborhood to send a message.  The latter
measure would allow for queuing, processing, and transmission delays as
well as extra delays due to the medium access control layer for collision
avoidance.  The first row of the table reflects that a read-all operation
is initiated by one node, followed by each of its $n-1$ neighbors 
sending a response.  A write-all operation potentially has 
the least resource cost of any operation, consisting of 
just one broadcast message; however to provide for reliability, 
an implementation of write-all may return acknowledgments from
each recipient of the broadcast back to operation's initiator.
For this reason, the number of message primitives is reported 
as ``1 or $n$'' in Table~\ref{fig:table-ops}.  A local 
transaction, as defined in \cite{D07} and called ``transact'' in
Table~\ref{fig:table-ops}, has a read set of $r$ nodes
and a write set of $w$ nodes.  The transaction is initiated with a 
broadcast, followed by a response from each node in the read set.
Then the transaction initiator transmits a broadcast to the
write set containing values to be written, and each member of 
the write set unicasts an acknowledgment to the initiator;  
the acknowledgments are needed so that the initiator can decide
whether to allow the transaction to commit or to broadcast a cancel message.
The read and write sets may overlap, with the worst case being 
$r=w=n-1$ (which would put the message cost of transact at $2n$). 
The LRW operation begins with a broadcast, followed
by each of the $n-1$ other nodes responding.  Since any value to 
be written is contained in the initial broadcast and responses
are collected by the LRW initiator, no additional 
round is needed to complete the operation.  

\begin{table}[t]
\begin{minipage}{\columnwidth}
\begin{center}
{\footnotesize
\begin{tabular}{|l|c|c|} \hline
\textsc{Operation} & \textsc{Messages} & \textsc{Rounds} \\ \hline
read-all & $n$ & 1 \\ \hline
write-all & 1 or $n$ & 1 \\ \hline
transact & $2+r+w$ & 2 \\ \hline
LRW & $n$ & 1 \\ \hline
\end{tabular}
}
\end{center}
\end{minipage}
\caption{\footnotesize Operation comparison.}
\label{fig:table-ops}
\end{table}

The measures of Table~\ref{fig:table-ops} are optimistic numbers
in two senses.  First, the measures are for transactions that succeed,
that is, they do not fail due to conflicts with concurrent transactions
or negative responses (an LRW operation would need to 
include a cancellation message if any neighbor response indicated
some unanticipated value).  Secondly, the table does not include 
commit messages for transaction or LRW operations.  This is because 
sensor node timing and clocking mechanisms enable commit to be 
time-triggered, that is, each node commits a transaction after sufficient
time has passed without receiving a cancellation message.   

\section{LRW Design Issues} \label{section:lrw-design}

\emph{Local Read-Write} (LRW) is an operation defined on variables of 
WSN nodes.  We assume that each node has the same set of 
variables\footnote{This assumption is not essential, but simplifies
the description.} that can be read and written by an LRW 
operation.  For variable $v$ and node $q$, let $v_q$ refer to 
$q$'s instance of $v$.  Each invocation of LRW specifies:
\itp{i} a function $f$ defined on a subset of node variables, 
\itp{ii} a subset of node variables to be written, and 
\itp{iii} a boolean function $g$.   Function $f$ can be 
computed at any node, and either returns a \emph{negative
response} value $\perp$ or returns a pair $(r,B)$, where $r$ is 
a value provided for computing $g$ and $B$ is a list of values
to be written to the variables specified in \itp{ii}.  
A \emph{nonlossy} LRW operation is defined 
with respect to an initiating node $p$ and $p$'s neighborhood
$N(p)$, consisting of three steps:
\bfp{1} for each node $q\in N(p)$, function $f$ is computed; 
\bfp{2} function $g$ is computed on the set of $r$-values 
$\{\,r_q\;|\;q\in N(p)\,\}$; and \bfp{3} if the result of $g$ is
\emph{true}, then $B_q$ is written to the write variables 
of $q$, for each node $q$.  A lossy LRW operation would allow,
in \bfp{1}--\bfp{3}, proper subsets of $N(p)$ to model the loss
of messages. We do not formally specify lossy LRW 
instances in this paper.     

Viewed from the application perspective, an LRW operation 
begins when initiating node $p$ invokes LRW and ends when 
$p$ receives a response from the LRW.   Between the invocation
and response, we assume that $p$ does not invoke another LRW
instance.  Thus the only source of concurrency in the system
is contention among LRW operations of different LRW initiators.
The behavior of a set of (possibly concurrent) 
LRW operations can be specified by a sequence, called an LRW 
history, which contains LRW invocations, contains results of $f$ and 
$g$ evaluations, assignments to variables, and 
contains responses to the LRW invocations.  
We omit details of the history formalization, which  
follow from standard techniques similar to the notation of 
transaction serializability.  A well-formed LRW history is one
in which every LRW invocation finds a matching response.  
A well-formed LRW history determines values for all variables.
Implementations of LRW or similar operations result in refined
histories, where between invocation and response, lower-level 
events (transmission, reception, message processing) occur.
Analysis of such operation histories, for implementations
of operations in Table~\ref{fig:table-ops}, can verify 
their atomicity properties.

The framework \cite{D07} uses terminology of transactions to 
describe local operations, including some ACID properties of 
transactions in databases. Atomicity of a transaction, which
is the all-or-none property, is due to two properties of the 
protocol.  First, in the WSN model, ordering transactions can 
be simple because message propagation latency is negligible.
If nodes $p$ and $p'$ concurrently initiate a transaction using
local broadcast, with $x,y\in N(p)$ and $x,y\in N(p')$, then 
$x$ and $y$ cannot receive broadcasts from $p$ and $p'$ in 
different order.  Second, all the writes of a transaction are
sandboxed and only actually written upon the event of 
transaction commit.  Consistency of transactions is ensured by 
conflict resolution.  If the transactions of $p$ and $p'$ 
conflict, say because they write to the same variable in 
node $x$, then one of the two transactions will be aborted 
(and possibly re\"initiated later).  Properties corresponding 
to atomicity and consistency can similarly be shown for 
LRW operations (and proved using LRW histories).  Transactions
of $p$ and $p'$ can be concurrent, even with neighbors 
$\{x,y\}$ in common, provided that they operate on distinct 
sets of variables (and more generally, if it can be shown that
the transactions have commutative semantics).  This observation
also holds for LRW operations.  

The problematic aspects of ACID properties for WSNs arise 
from platform limitations and unreliable message transport.  
The possibility of message loss implies, for example, that 
a commit message or a cancellation message could be lost. 
It is well-known that no acknowledgment protocol can guarantee
that all neighbors of a transaction initiator will receive
a commit or cancellation message, even if it is retransmitted
some number of times \cite{G78,AR94}.  However, the probability of a  
communication loss can be reduced if messages are retransmitted, 
and retransmission may be a practical strategy to improve 
reliability for WSNs (in effect, retransmission is an approximation
to eventually correct message delivery).  A protocol optimization 
for transaction or LRW operations is to replace a commit or 
cancellation message with timeout-driven
activation.  The design choice of \cite{D07} and in this paper is 
to let commit be timeout-driven:  if, after some fixed time
period, a node does not receive any cancellation message from
the LRW initiator, then variables writes are committed.  
An alternative design choice would be to let cancellation be 
the timeout-triggered default, however this choice would shift
the balance of power usage (because messages consume power)
to commit, and for most applications and typical WSN workloads,
one would expect most LRW operations to be committed.

A limitation of several current WSN message protocols is packet
payload size.  For the platform used in our experiments, the 
payload is 28 bytes, which limits how much can be specified in
an LRW operation based on a single broadcast.  Scaling LRW to 
larger data amounts would require fragmentation of LRW message
fields over multiple broadcasts.  Some WSN platforms may not 
support native local broadcast;  there, ordering LRW operations 
by the instant of reception would not be reliable.  However 
LRW operations can also be ordered by timestamp, if the WSN 
has synchronized clocks.  With synchronized clocks, LRW operations
can be grouped by slotted time intervals.  In a slotted time 
protocol, initiators wait until the beginning of a slot before 
transmitting an LRW operation message;  when a neighbor receives
an LRW message, it delays sending a response until the end of 
the current slot, in order to collect all LRW operations, order
them, and sort out conflicts.  A reason to consider using 
unicast, rather than broadcast of the initial LRW message, is 
to improve scheduling efficiency of responses from neighbors.  
The CC2420 radio chip has a feature for immediate acknowledgment of  
unicast messages, and this feature is not available for local broadcast.

In the discussion above, we have treated $N(p)$ as a constant, 
supposing the neighborhood of $p$ to be fixed in the WSN.  The experience
of many researchers is that, even for a static WSN, radio properties
are dynamic:  the set of stable, bidirectional links defining neighborhoods  
evolves.  Therefore the design of an LRW protocol should plan for 
dynamic neighborhoods.  If an LRW operation fails because the initiator
did not collect responses from every neighbor (this would depend on 
the definition of $g$), it could be that the neighborhood has changed. 
In this case, subsequently submitting the LRW operation would
use the new neighborhood.

\section{LRW Operation Comparison} \label{section:lrw-theory}

Table~\ref{fig:table-ops} does not 
compare expressive, or computing power of different neighborhood
operations.  In the table, transact consumes most resource, but 
transact is more powerful than any other:  in one operation, 
a function of neighborhood values can be computed and written
to several nodes.  An LRW operation is strictly less powerful because 
any value written must be prescribed, before the operation is 
invoked, rather than computing the value to write during the operation.  

One technique to compare operation power is to examine protocols
that use only that operation to solve some classic problem, such 
as consensus.  If one operation type enables consensus to be solved
whereas another operation does not, then the former operation is
more powerful (with respect to consensus) than the latter.  
Briefly, a consensus protocol begins with each node having an 
input value and a decision variable, which
can be written at most once.  The input is not 
in any variable, that is, input values cannot directly be
viewed by any of the operations of Table~\ref{fig:table-ops};
an early step in any consensus protocol is to share the input
with other nodes.  The initial value of the decision 
is some constant $\omega$ not equal to any node's input.  
Consensus protocols must satisfy 
three properties: validity, agreement, and termination. 
The termination property is that every node eventually writes to
its decision variable, regardless of the progress or failure
of other nodes; agreement requires that no two nodes write
different decision values;  validity requires that any 
decision written be the input of some node.  
The difficulty of consensus lies in the timing of nodes 
participating in the protocol.  If some node $p$ is very slow to 
engage in the protocol, then other nodes will need to decide without
knowing $p$'s input.   Although synchronous timing is implicit
in the implementation of operations such as LRW, software at the
application layer may be asynchronous, hence the timing of 
applications using LRW can be unpredictable.

For the following results, we assume communications are nonlossy 
and do not fail due to contention conflicts.  Also, neighborhoods
are static and definitions of neighborhood are consistent, that
is, if $q\in N(p)$ then $p\in N(q)$.  The following shows that
read-all is insufficient to solve the consensus problem.

\begin{lemma} \label{lemma-noread-consensus} 
Consensus using only read-all operations is impossible.
\end{lemma}
\begin{proof}  The proof repeats standard arguments \cite{H91} based
on finding a contradiction in a constructed execution.  Suppose 
consensus is possible, and that nodes $p$ and $q$ are neighbors
with inputs 0 and 1 respectively.  If $p$ (or $q$) waits long enough
to expose its input value, then the other node may take sufficiently
many steps so that it is forced, by the termination property, to 
decide; because the other's input is unknown, it will 
decide in favor of its own input.  Thus the
initial state for the consensus protocol is \emph{multivalent}, that is,
there exist two possible executions leading to different decisions. 
A state is \emph{univalent} if all possible executions following that
state can only lead to one decision (in effect, the decision has already
been chosen, even if not presently in a decision variable).  Executions
consist of an interleaving, of atomic steps from some node in the 
neighborhood, where a step is either a read-all operation, some 
local calculation, or writing to some variable(s).  If $p$ writes
to a variable $v$, and the next step in the execution is a 
read-all for $v$ by $q$, then $q$ obtains the value $p$ wrote to $v$.
\par
Let $\sigma$ be the last multivalent state in an execution (the
termination property implies $\sigma$ exists).  There are at least 
two possible continuations from $\sigma$ leading to different decisions,
by definition.  Such continuations necessarily begin with steps of 
different nodes.  We consider different cases for the first step
by $p$ and $q$ with respect to continuations.  Note that if $p$
steps first after $\sigma$, then the valency is different than 
would be if $q$ steps first (otherwise $\sigma$ is not multivalent).
If the first step by $p$ is a local calculation or a read-all
operation, then the occurrence of that step is undetectable by $q$. 
This contradicts the assumption that $p$'s first step after $\sigma$ 
results in a univalent state.  If the 
first step by $p$ writes to a variable, then it 
cannot be that $q$'s first step writes to a variable, because
these two steps commute, which would contradict the differing valency of
these two steps.  Therefore, the essential case to examine is where
$p$'s first step writes to a variable and $q$'s first step 
is a read-all operation.  If $p$ steps first and then sleeps while
$q$ runs long enough to decide, the valency will follow from $p$'s
write of a variable;  the same valency is obtained if $q$
makes no steps while $p$ runs long enough to decide.  However, $p$
cannot detect whether or not $q$ has performed a read-all, hence
if $q$ steps first, then sleeps, with $p$ running long enough to decide,
$p$ must decide as if $q$ took no steps, which contradicts the 
supposed valency of $q$'s read-all operation.  Thus in any case, 
the transition from multivalency to univalency can be prevented in
some possible execution.
\end{proof}

Although read-all doesn't provide a solution to consensus, an 
enhanced form of read-all, called \emph{read-all-write}, does
allow for a solution.  In a read-all-write operation, a node 
atomically reads values from all neighbors and writes some 
function of the result to a variable.  If $p$ and $q$
invoke read-all-write at nearly the same instant, then atomicity
guarantees that one operation will precede the other.  Thereby,
if $p$'s read-all-write occurs first, then the variable
written by $p$ will be visible in $q$'s read-all-write.  Thus
$q$ can detect that $p$'s operation preceded $q$'s, and the 
decision value for both nodes can be the input of $p$.  
A read-all operation is ``lighter weight'' compared to a
read-all-write operation, which must constrain concurrency
to guarantee atomicity.  We are not aware of WSN research on 
neighborhood read-all-write.  Presumably the transactional methods, 
say of \cite{MN98} or \cite{D07} could be used to implement read-all-write. 

Unlike read-all, the write-all operation can be used to 
solve consensus in particular cases.  The following first 
identifies a negative case, where write-all is insufficient; 
afterward we discuss a case where a consensus protocol
uses write-all.  
\begin{lemma} \label{lemma-nowrite-consensus} 
Consensus using only single-variable write-all 
operations is impossible.
\end{lemma}
\begin{proof}
The proof is similar to that for Lemma \ref{lemma-noread-consensus}.
Here, each node obtains values of other nodes only by locally reading 
variables that have been assigned by a write-all operation.
Let $\sigma$ be the last multivalent state in an execution, and 
suppose the next steps of $p$ and $q$ are write-all operations. 
If these steps write to different variables, then the steps
commute and a valency contradiction is obtained.  When $N(p)=\{q\}$
the steps of $p$ and $q$ write to different variables because
write-all operations assign to variables of other nodes.  
One case where two steps write to the same variable, 
is that $p$'s first step writes to its variable $v_p$ and $q$'s
first step is a write-all to $v_p$.  Suppose the valency of 
$p$ taking the first step is 1, and the valency of $q$ taking the
first step is 0.  If $q$ takes the first step and $p$ sleeps long
enough for $q$ to decide, the decision is 0.  If $p$ takes the
first step and then sleeps while $q$ runs long enough to decide,
the decision will still be 0, because $q$ overwrote what $p$ had
written, and thus $q$ is not influenced by $p$'s initial step.  
This contradicts the assumption that $p$'s first step results 
in a univalent state with valency 1.
\end{proof}
The write-all operation of Table~\ref{fig:table-ops} does not
include any local variable as a write target (that is,
$p\not\in N(p)$).  An extension to write-all would be to include
$v_p$ in $p$'s write-all of variable $v$.  This extension alone turns
out not to help in solving consensus, however the inclusion of
$v_p$ together with allowing multiple variables to be written
does enable a consensus protocol.  Suppose $N(p)=\{q\}$ and 
three variables $u$, $v$, $w$ are initially $\omega$.  Let 
$p$'s operation write its input value to $u$ and 
to $v$;  and let $q$'s operation write its input
value to $v$ and to $w$.  Whichever node has the first write-all 
operation forces the decision value to be its input. 
In an execution with differing inputs such that $p$ invokes 
the first write-all and $q$ sleeps, $p$ will detect that 
it has the first operation, because $w=\omega$;  and if 
$q$ does not sleep, $p$ may detect that $w\neq\omega$, 
however then $v$ does not contain $p$'s input, and so 
$p$ detects that its write-all occurred first ($q$ will
get the decision from $v$ in that case).  

It is not difficult to show that LRW or transact suffice
to solve consensus, because it is simple for a node to 
record a decision value that is not overwritten by any
subsequent operation.  The LRW operation is a 
lighter weight primitive than write-all, 
which deals with more variables than
LRW when used for for consensus.  

When operations have equivalent solvability power, 
then may also be compared by the time 
required or the number of operations used in a solution. 
Intuitively an LRW operation does more work per operation 
than either read-all or write-all operations, and all 
of these have 1-round (optimistic) time complexity.  

\section{Implementation} \label{section:lrw-experiments}

The previous sections of the paper motivate LRW operations and sketch,
at a high level, how such an operation could be implemented in a WSN.  
To confirm the feasibility of LRW on a current sensor network platform,
this section reports results from simple experiments on some small mote networks.
Section \ref{section:lrw-design} exposes general design issues for an
implementation, whereas the experimental implementation must contend with
low-level design considerations.  For example, the table in 
Figure \ref{fig:table-ops} reports optimistic message counts for LRW, but
our experiments consider failures in message delivery.  Operation duration 
and throughput is affected by thresholds for message transit time 
and expected number of retransmissions;  such factors are determined from 
experiments.  For instance, the duration of an LRW operation can be reduced 
by setting smaller time limits and lowering the number of retries for
lost messages;  this will allow more LRW operations to be executed, at 
the cost of reliability.  Figures presented below show effects of such
tuning decisions, for the case of an LRW operation run in isolation and also
for the case of an LRW contending with other operations.

\subsection{Experimental Platform}

\begin{figure}[ht]
\framebox{
\begin{minipage}{\columnwidth}
{\footnotesize
\begin{tabbing} 
xxx \= xxx \= xxx \= xxx \= \kill
\textbf{LRW-initiate($p$):} \\
\> $\textit{mode}\leftarrow\textit{active}$, $S\leftarrow\{\;\}$  \\
\> \textit{start} $\textsf{T}_{Timeout}$, $\textsf{T}_{Commit}$ \\
\> \textbf{broadcast}( $\textrm{initMsg}_p(\textsf{T}_{Commit})$ ) \\
\> \textit{start} $\textsf{T}_{Response}$ \\
\> \textbf{while} ($\textit{mode}=\textit{active}$) \\ 
\>\> \textbf{receive}($\textrm{rejectMsg}_q$) : \\
\>\>\> \textit{mode} $\leftarrow$ \textit{cancel} \\
\>\> \textbf{receive}($m=\textrm{acceptMsg}_q$) : \\
\>\>\> $S \leftarrow S \cup \{ sender(m) \}$ \\ 
\>\>\> \textbf{if} $|S|=|N(p)|$ \\ 
\>\>\>\> \textit{stop Timers}; \textbf{return} \textit{success} \\ 
\>\> $\textsf{T}_{Response}$ \textit{expires} : \\
\>\>\> \textbf{broadcast}($\textrm{initMsg}_p(\textsf{T}_{Commit})$) \\
\>\>\> \textit{restart} $\textsf{T}_{Response}$ \\
\>\> $\textsf{T}_{Timeout}$ \textit{expires} : \\
\>\>\> $\textit{mode}\leftarrow\textit{abort}$,
	$S\leftarrow \{\;\}$ \\
\>\>\> \textbf{broadcast}( $\textrm{abortMsg}_p$ ) \\
\>\>\> \textit{start} $\textsf{T}_{Response}$, $\textsf{T}_{Timeout}$ \\ 
\> \textbf{while} ($\textit{mode}=\textit{abort}$) \\
\>\> \textbf{receive}($m=\textit{abortAck}_q$) : \\
\>\>\> $S \leftarrow S \cup \{ m \}$ \\
\>\>\> \textbf{if} $|S|=|N(p)|$ \\ 
\>\>\>\> \textit{cancel Timers}, \textbf{return} \textit{canceled} \\ 
\>\> $\textsf{T}_{Response}$ \textit{expires} : \\
\>\>\> \textbf{broadcast}( $\textrm{abortMsg}_p$ ) \\
\>\>\> \textit{start} $\textsf{T}_{Response}$ \\ 
\>\> $\textsf{T}_{Timeout}$ \textit{expires}  : \\
\>\>\> \textit{stop Timers}; \textbf{return} \textit{failed} 
\end{tabbing}
}
\end{minipage}
}
\caption{LRW for Initiator $p$}
\label{newfig-init}
\end{figure}

Our implementation and experiments were written in the NesC language for
the TinyOS (version 2) operating system, running on Telosb \cite{PSC05} 
and MicaZ motes (both platforms use the same radio chip, CC2420).  
Rather than a full 
implementation of LRW, we used a simpler protocol that ignored the case
of application-triggered operation failures (such as one neighbor having
a value that cancels the LRW operation);  thus all our experiments 
consider only cases of successful LRW operations, except where an 
operation is rejected due to concurrency.  The implementation is built 
on several services:  a MAC-layer radio stack transmits and delivers
packets, also inserting random delays (typically between 3ms and 12ms)
to avoid collision with other transmissions;  a neighborhood service 
determines the effective set of a node's neighbors (for which there
is currently bidirectional communication);  a clock synchronization service
aligns the timers of nodes, which facilitates experiments that induce
concurrent LRW operations in a controlled way.  Our largest experiments
used 31 MicaZ motes, and due to proximity, we artificially constrained
each node to have at most six neighbors (essentially, this is topology
control).  Our implementation employed three countdown timers, two for
message delivery and acknowledgment, and a third to limit the total duration
of the LRW operation.  We used two timers for message delivery, to 
make the distinction between \itp{i} time for message transmission 
and receiving a response message, and \itp{ii} the time for successful
transmission and response from all neighbors, including retries of \itp{i}.  
A technical reason to use \itp{ii} instead of a retry counter is that
the MAC layer's timing is randomized, and our design goal was to 
implement a protocol with known thresholds for the LRW operation.  Should
the timer for \itp{ii} expire, then the LRW operation's initiator aborts
the operation and transmits abort commands to its neighborhood.  The 
third timer is for operation commit:  if a neighbor does not receive an
abort message and the commit timer expires, then the result of the 
LRW is committed.

\subsection{Three Outcomes for LRW}

We say that an LRW operation has three possible outcomes: 
It is considered a \emph{Success} if the LRW is accepted by all neighbors; 
the operation is considered \emph{Canceled} if an abort message 
(which may have become necessary for a number of reasons) is responded to by 
all neighbors; and it is considered \emph{Failed} if at least one neighbor 
does not respond to the abort message.  Each LRW operation returns to 
the application, which invoked LRW, one of these three outcomes.
The first two outcomes, \textit{Success} 
and \textit{Cancel}, are within the intended behavior of the protocol (in the
terminology of transactions, the result satisfies Atomicity and Consistency
criteria).  The final outcome, \textit{Failed}, represents a failure of 
communication, a sensor node crash, or (silent) neighborhood reconfiguration.
For a failed outcome, it is uncertain whether or not all neighbors received
and committed the LRW operation (possibly, the initiator attempted to 
cancel the operation, but not all neighbors acknowledged the cancel request
within the allowed timeout period).  For a set of LRW operations, \emph{reliability} 
is the percentage of non-failed LRW operations, that is, operations which 
respond by success or cancel.

\begin{figure}[ht]
\begin{center}
\framebox{
\begin{minipage}{\columnwidth}
{\footnotesize
\begin{tabbing} 
xxx \= xxx \= xxx \= xxx \= \kill
\textbf{LRW-neighbor($q$):} \\
\> \textbf{initially:} $\textit{mode}=\textit{idle}$ \\
\> \textbf{while} ($\textit{mode}=\textit{engaged}_p$) \\
\>\> \textbf{receive}( $\textrm{abortMsg}_p$ ) : \\
\>\>\> \textit{stop} $\textsf{T}_{Commit}$ \\
\>\>\> \textbf{send}( $p, \textrm{abortAck}_q$ ) \\ 
\>\>\> $\textit{mode} \leftarrow \textit{idle}$  \\
\>\> \textbf{receive}( $m=\textrm{initMsg}_p(t)$ ) : \\
\>\>\> \textbf{send}( $p$, $\textrm{acceptMsg}_q$ ) \\
\>\> \textbf{receive}( $\textrm{initMsg}_r \;\wedge\; r\neq p$ ) : \\
\>\>\> \textbf{send}( $r, \textrm{rejectMsg}_q$ ) \\
\>\> $\textsf{T}_{Commit}$ \textit{expires} : \\
\>\>\> $q$ \textit{commits LRW} \\
\>\>\> $\textit{mode} \leftarrow \textit{idle}$  \\
\> \textbf{while} ($\textit{mode}=\textit{idle}$) \\
\>\> \textbf{receive}( $\textrm{abortMsg}_r$ ) : \\
\>\>\> \textbf{send}( $r, \textrm{abortAck}_q$ ) \\ 
\>\> \textbf{receive}( $m=\textrm{initMsg}_p(t)$ ) : \\
\>\>\> \textit{start} $\textsf{T}_{Commit}\leftarrow t$ \\
\>\>\> $\textit{mode} \leftarrow \textit{engaged}_p$ \\
\>\>\> \textbf{send}( $p$, $\textrm{acceptMsg}_q$ ) \\
\end{tabbing}
}
\end{minipage}
}
\end{center}
\caption{LRW for Neighbor $q$}
\label{newfig-neighbor}
\end{figure}

\subsection{Protocol}

In the following, $p$ refers to an initiator and 
$\{q_0, q_1, ..., q_k\} = N(p)$ refers to the neighbors of $p$.
Figures \ref{newfig-init} and \ref{newfig-neighbor} 
contain a high-level description
of the LRW implementation.  Five message types are used in the protocol,
\textit{initMsg}, \textit{acceptMsg}, \textit{rejectMsg}, 
\textit{abortMsg}, and \textit{abortAck}.  Three event types drive
the protocol:  invocation of an LRW operation, message arrival, and
timer expiration.  The protocol timers are denoted $\textsf{T}_x$ where
$x\in\{\textit{Response,Timeout,Commit}\}$.  Each timer starts with 
some positive value and decreases to zero, whereat an expiration event
occurs.   The initiator begins by starting three timers, however only 
$\textsf{T}_{Response}$ and $\textsf{T}_{Timeout}$ have expiration events 
shown in the initiator protocol; $\textsf{T}_{Commit}$ 
provides a time stamp for the LRW 
operation, used by neighbors. (Note that $\textsf{T}_{Commit}$ could be
significant to the initiator as a safeguard, to reject any 
application invocation of LRW-initiate while a current invocation
is already in progress;  $\textsf{T}_{Commit}$ could also trigger
commit at the initiator itself, but to simplify the presentation
we suppress such details.)

The first message broadcast by the initiator 
is the \textit{initMsg}, containing the current $\textsf{T}_{Commit}$ value 
and other fields relevant to the LRW operation.  The initiator waits for all 
neighbors to reply with \textit{acceptMsg} (if $\textsf{T}_{Response}$ 
expires, then the initiator again broadcasts an \textit{initMsg}).  To simplify
the description, the case where $\textsf{T}_{Commit}$ expires (which could
occur if the initiator retries a broadcast too many times) is not shown.  
Also, the presentation omits the case where a collection of \textit{acceptMsg} 
values might trigger some application-specific abort of the LRW operation.  

For the mote implementation, we also included a list of neighbors in the 
\textit{initMsg}:  this list named the neighbors for which the initiator 
had not yet received a \textit{acceptMsg}.  This small optimization reduced
the overhead after retransmitting an \textit{initMsg},  by avoiding 
a needless resending of \textit{acceptMsg} (usually for the majority of 
neighbors).

It is important to note that even if \textbf{LRW-initiate} 
returns \textit{Success}, $\textsf{T}_{Commit}$ may not have expired, 
and thus the LRW has not been committed. For this reason $p$ must not be 
permitted to begin a new LRW operation until $\textsf{T}_{Commit}$ expires. 
In our trials, the frequency of LRW operations was small enough to ensure 
that the previous LRW was committed or aborted before a new LRW was started, 
but any implementation of this protocol would have to address the 
possibility of such an occurrence. Contrary to this, if the LRW operation 
is aborted, a new LRW may be initiated immediately.

If the LRW fails due to a neighbor $q_i$, then since at least one attempt 
has been made by the initiator $p$ to cancel the operation, we know that 
one of the following is true: \textit{(1)} $q_i$ did not receive the 
initiation message; \textit{(2)} $q_i$ received the initiation message, 
but not the abort message; or \textit{(3)} $p$ did not receive $q_i$'s 
response to the abort message. In cases \textit{(1)} and \textit{(3)} 
the LRW is not committed, and thus consistency is maintained. In Case 
\textit{(2)} however, the LRW is committed, and consistency is lost.

\subsection{Metrics} \label{sec:metrics}
We evaluated the LRW implementation with respect to time and reliability. 
For experiments, we considered different platforms, different topologies, 
different low-level choices for communication, whether 
operations are run in isolation or operations run are under contention, as well as different
settings for $\textsf{T}_{Timeout}$ and $\textsf{T}_{Commit}$.  
The total time allotted to an LRW operation is $\textsf{T}_{Commit}$, 
however under ideal circumstances, 
all messages are sent, delivered, and acknowledged in a time possibly 
much less than $\textsf{T}_{Commit}$:  we refer to the time 
needed for one LRW operation to complete under these circumstances as 
the \emph{optimistic duration} of an LRW operation. Measuring the optimistic duration is interesting:
if the gap between the optimistic duration and $\textsf{T}_{Commit}$ is large, and if close to 
ideal circumstances are common, then $\textsf{T}_{Commit}$ may be decreased. A smaller
value of $\textsf{T}_{Commit}$ would allow an application to submit more LRW operations,
that is, the throughput of LRW operations could be higher if the operation 
interval is shorter.  However, decreasing $\textsf{T}_{Timeout}$ and $\textsf{T}_{Commit}$  
may also decrease the frequency of success, because fewer message retries will
be attempted and late-arriving messages could be discarded.  A less reliable
implementation impacts the application, which may or may not retry an unsuccessful 
LRW operation.  In view of the two unsuccessful outcome possibilities for 
LRW-initiate, \textit{canceled} and \textit{failed}, the application should
decide whether or not to retry a canceled operation. Reducing the 
frequency of \textit{failed} cases can be handled within the LRW implementation.

\begin{figure}[ht]
\begin{minipage}{\textwidth}
\psset{xunit=5mm,yunit=0.3mm}
\psset{yticksize=0 5cm}
\begin{pspicture}(-3,-50)(15,150) 
\savedata{\unicast}[(1,13)(2,25)(3,36)(4,48)(5,59)(6,71)(7,85)(8,96)(9,109)(10,121)(11,131)] 
\savedata{\multicast}[(1,24)(2,28)(3,32)(4,39)(5,48)(6,57)(7,63)(8,76)(9,84)(10,91)(11,100)] 
\psframe[fillstyle=solid,fillcolor=white,linestyle=solid](1,0)(11,140)
\psaxes[Dy=20,Ox=1,Oy=0](1,0)(11,140)
\rput(6.3,-35){\textsf{Neighborhood Size}}
\rput(-1.7,70){\rotatebox{90}{\textsf{Milliseconds}}}
\dataplot[plotstyle=dots,linecolor=red,dotstyle=diamond*]{\unicast}
\dataplot[plotstyle=line,linecolor=red]{\unicast}
\dataplot[plotstyle=dots,linecolor=blue,dotstyle=square*]{\multicast}
\dataplot[plotstyle=line,linecolor=blue]{\multicast}

\pspolygon[fillcolor=white,fillstyle=solid](1.2,135)(1.2,102)(5.4,102)(5.4,135)
\psdots[linecolor=blue,dotstyle=square*,dotsize=5pt](1.6,128)
\rput(3.6,128){\color{black} Broadcast}
\psdots[linecolor=red,dotstyle=diamond*,dotsize=5pt](1.6,110)
\rput(3.2,110){\color{black} Unicast}
\end{pspicture} 
\end{minipage}
\caption{Broadcast vs Unicast with Ack}
\label{fig:broad_uni_nocont}
\end{figure}
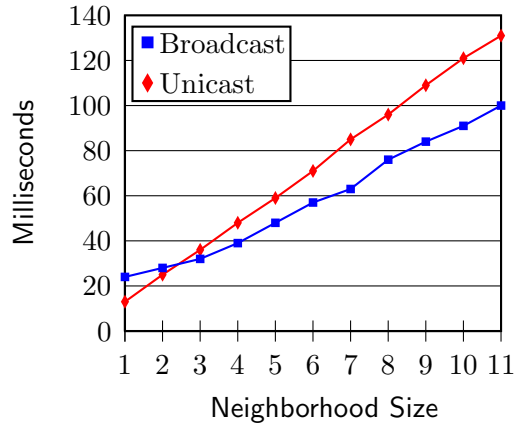

To measure the implementation under contention, we used the synchronized clock
service \cite{Ia08} 
to arrange that a set of nodes simultaneously invoke LRW-initiate.  
Within an experiment, a set of LRW operations started simultaneously is called a 
\emph{series};  the duration of that series is defined to be the maximum optimistic 
duration of any LRW operation in the set. We say that a series \emph{failed} 
if at least one operation in it \textit{failed}, and that 
it \emph{succeeded} otherwise. We define the reliability of a set of 
series as the percentage of successful operations. Our experiments show factors that 
influence a series duration and reliability.

\subsection{Experiments}

\textbf{Communication Primitive.}  Each \textit{initMsg} from an initiator should
be acknowledged, either with an \textit{acceptMsg} or a \textit{rejectMsg}, by each neighbor.  
The CC2420 radio chip offers a hardware-level acknowledgment feature, which
enables the receiver of a unicast message to send an ack frame immediately,
without the usual MAC delay (to avoid collision).  Though this feature does not
presently provide a payload area within the acknowledgment frame, and hence
is not sufficient for LRW purposes, we nonetheless tested how well using 
unicast (with the hardware acknowledgments) compared to using neighborhood broadcast.  
Using unicast requires more transmission operations by
the initiator (one per neighbor), but also saves time communicating acknowledgments. 
Figure \ref{fig:broad_uni_nocont} displays results of an 
experiment conducted on Telosb motes, where each data point in the graph
is the mean of approximately 250 successful operations.  
Due to a generous timeout, no operations were aborted or failed. In this experiment, there is only one initiator, 
and the number of neighbors varied, shown on the x-axis;
the optimistic duration is shown on the y-axis.  Because this experiment showed
the superiority of using broadcast with denser networks (and the current 
unicast with hardware ack is deficient for our purposes), we used the neighborhood
broadcast primitive in all subsequent experiments.  Figure \ref{fig:broad_uni_nocont}
also suggests a starting point for testing different values of $\textsf{T}_{Timeout}$ at 
different neighborhood sizes.  For example, given an initiator with six neighbors, 
50ms might be a starting point for an experiment testing reliability.

\paragraph{Reliability and Timeout}

Recall from Section \ref{sec:metrics} that the duration 
of the $\textsf{T}_{Commit}$ timer is a significant factor in the 
throughput of the protocol. If the protocol is expected to rapidly execute 
several LRW operations, there is a significant incentive to keep 
the $\textsf{T}_{Commit}$ timer as short as possible. This however 
carries with it another set of challenges:  note that the 
execution of the LRW operation must be contained entirely 
within the timespan of $\textsf{T}_{Commit}$. Furthermore, 
observe from Figure \ref{newfig-init} that $\textsf{T}_{Timeout}$ 
performs a slightly different function depending on the current 
mode of the initiator $p$; if it is in \textit{active} mode the 
timer is used to enter the initiator into \textit{abort} mode, 
and if it is in \textit{abort} mode, the timer signals when the 
operation is to be declared as \textit{failed}. Since the 
$\textsf{T}_{Commit}$ timer must be of sufficient duration to allow for 
both of these eventualities it follows that the length of 
$\textsf{T}_{Commit}$ must be at least twice that of $\textsf{T}_{Timeout}$.

\begin{table}[ht]
\begin{minipage}{\columnwidth}
\begin{center}
{\footnotesize
\begin{tabular}{|r|l|}
\hline
Timeout & Reliability\\
\hline
50ms & 46.64\% \\ 
\hline
75ms & 96.87 \% \\ 
\hline
100ms & 100 \% \\ 
\hline
\end{tabular}
}
\end{center}
\end{minipage}
\caption{Reliability without contention.}
\label{tab:rel_nocont}
\end{table}

Due to the above reasoning, many of our experiments were focused 
on studying the effect that reducing the duration 
of $\textsf{T}_{Timeout}$ would have on the duration and reliability 
of both operations and series. We performed two sets of trials.  The 
first set was executed in a simple network containing one initiator 
with a static neighborhood of size six. In these experiments we used TelosB motes.

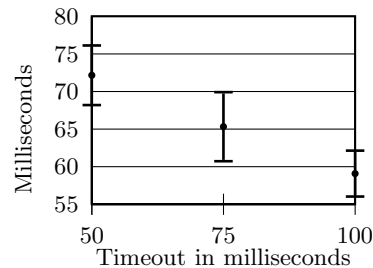
\begin{figure}[ht]
\begin{minipage}{\columnwidth}
\readdata{\Data}{rel_cont.csv}
\psset{dotscale=0.75}%
\psset{xunit=0.07cm,yunit=.1cm}
\psset{yticksize=0 3.5cm}
\begin{center}
\begin{pspicture}(45,45)(100,80)

   {\footnotesize
 \psframe[fillstyle=solid,fillcolor=white,linestyle=solid](50,55)(100,80)
 \psaxes[Dy=5,Ox=50,Dx=25,Oy=55](50,55)(100,80)
 \rput{90}(37,65){Milliseconds}
 \rput(75,48){Timeout in milliseconds}
 \def\DoCoordinate#1#2{\psdot(#1,#2)}%
 \GetCoordinates{\Data}
 }
\end{pspicture}
\end{center}
\end{minipage}
\caption{Average LRW duration with 95\% confidence interval for varying timeout.}
\label{fig:lrw_dur_conf}
\end{figure} 

The second set introduced contention in the form of several initiators being present 
in the network at the same time. We also employed pre\"existing clock synchronization 
and neighborhood services. Note however that, as was mentioned in Section 
\ref{section:lrw-design}, even for a static WSN network the neighborhood of 
a single mote is often dynamic. The network in this case consisted of 31 MicaZ motes, 
labeled as $v_1, v_2, ..., v_{31}$. A mote $v_i$ would be considered an initiator 
if and only if $i \,\,\, mod(6) \equiv 1$ (thus the motes $v_1, v_7, v_{13}, 
v_{19}, v_{25}$ and $v_{31}$ were initiators). In order to control the topology 
of the network, we limited the potential neighborhood of an initiator $v_i$ 
such that $N(v_i) \subset \{v_{i-3}, v_{i-2}, v_{i-1}, v_{i+1}, v_{i+2}, v_{i+3}\}$. 
Observe that because of this, any two coordinators $v_i$ and $v_i+6$ 
will have overlapping potential neighborhood (due to using a dynamic neighborhood 
service, the actual neighborhood varies).

\begin{figure}[ht]
\begin{minipage}{\columnwidth}
\psset{xunit=0.79cm,yunit=.06cm}
\psset{yticksize=0 5.53cm}
\def\psvlabel#1{#1\,\%}
\begin{pspicture}(-2,-18)(7,60)
	{\footnotesize 
	\readpsbardata[header=true]{\data}{avg_time_nocont.csv}
  \psframe[fillstyle=solid,fillcolor=white,linestyle=solid](0,0)(7,60)
  \psaxes[labels=y,labels=y,Dx=1,Dy=10](0,0)(7,60)
  \psbarchart[barstyle={blue,green,red},barlabelrot=-27.5]{\data}
  }
  
  \pspolygon[fillcolor=white,fillstyle=solid](4.7,58)(4.7,35)(6.8,35)(6.8,58)
  
  {\footnotesize
  \rput{90}(-1.4,30){LRW operations}
  \rput(3.5,-18){Time interval in milliseconds}
  
  \psdots[linecolor=blue,dotstyle=square*,dotsize=8pt](5.0,55)
	\rput(5.9, 55){\color{black}50ms}
	
	\psdots[linecolor=green,dotstyle=square*,dotsize=8pt](5.0,47.5)
	\rput(5.9,47.5){\color{black}75ms}
	
  \psdots[linecolor=red,dotstyle=square*,dotsize=8pt](5.0,40)
	\rput(5.9,40){\color{black}100ms}
  }
\end{pspicture}
\end{minipage}
\caption{Duration distribution for varying timeout.}
\label{fig:nocont_dist}
\end{figure}
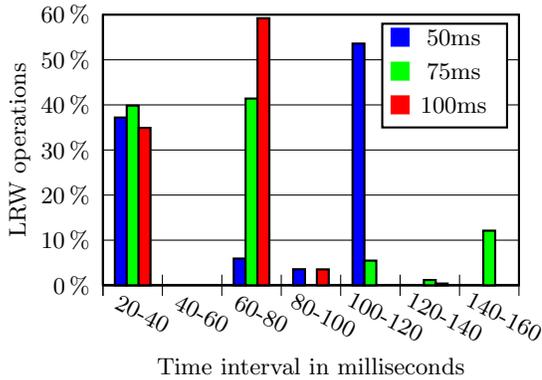

\paragraph{}
Our initial experiments were intended to evaluate the optimistic 
duration of the protocol. The implementation we used for these trials 
did not make use of clock synchronization or neighborhood services. 
Each trial consisted of approximately 250 LRW operations and the 
network contained only a single initiator with a neighborhood of size six.

As was previously mentioned, the duration of $\textsf{T}_{Timeout}$ is a 
major factor in the throughput. Thus we ran several experiments where we 
varied $\textsf{T}_{Timeout}$, and in each case noted not only the duration 
of each operation, but also the overall reliability. As was suggesting in the 
start of this section, we began with $\textsf{T}_{Timeout} = 50$ms, and 
incremented this in steps of 25ms until we achieved 100\% reliability. 
Table \ref{tab:rel_nocont} shows the reliability of each trial, 
and Figure \ref{fig:lrw_dur_conf} shows the average duration of 
the LRW operations with the 95\% confidence interval.

\begin{figure}[ht]
\begin{minipage}{\columnwidth}
\psset{xunit=0.15cm,yunit=0.5cm}
\psset{xticksize=0 8cm}
\def\pshlabel#1{#1\,\%}
\begin{center}
\begin{pspicture}(-2,-2)(30,16)
	{\footnotesize
	\psframe[fillstyle=solid,fillcolor=white,linestyle=solid](0,0)(30,16)
	\psaxes[ticks=x,labels=x,Ox=0,Oy=0,Dx=5,Dy=1](0,0)(30,16)
	\readpsbardata[header=true]{\data}{op_time_int.csv}
	\psbarchart[barstyle={green,red,blue,black},orientation=horizontal]{\data}
  \rput(14.5,-1.6){LRW operations}
  \rput{90}(-11,8){Time intervals in milliseconds}
  
  \pspolygon[fillcolor=white,fillstyle=solid](12,4)(12,0.5)(28,0.5)(28,4)
  
  \psdots[linecolor=black,dotstyle=square*,dotsize=8pt](14,3.5)
	\rput(20,3.5){\color{black}{\small 100ms}}
	
	\psdots[linecolor=blue,dotstyle=square*,dotsize=8pt](14,2.667)
	\rput(20,2.667){\color{black}{\small 150ms}}
	
  \psdots[linecolor=red,dotstyle=square*,dotsize=8pt](14,1.833)
	\rput(20,1.833){\color{black}{\small 250ms}}
	
  \psdots[linecolor=green,dotstyle=square*,dotsize=8pt](14,1)
	\rput(20,1){\color{black}{\small 350ms}}
	}
\end{pspicture}
\end{center}
\end{minipage}
\caption{Distribution of operation duration.}
\label{fig:cont_op_dur_dist}
\end{figure}
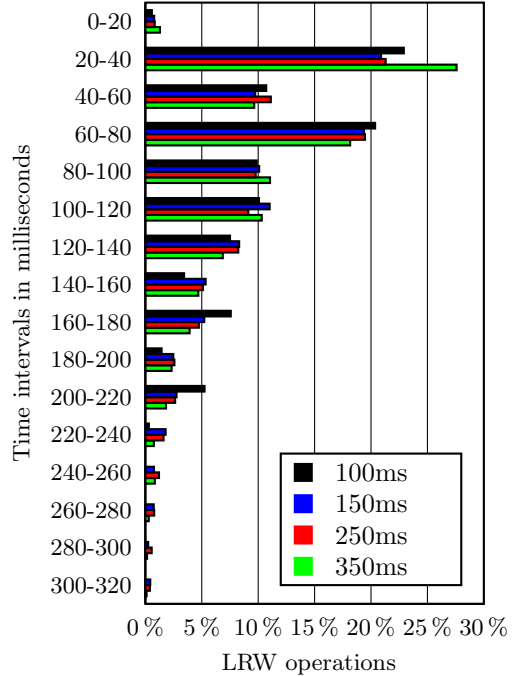

As seen in Table \ref{tab:rel_nocont}, with a 50ms timeout value the 
reliability was less than 50\%, but it increased to approximately 97\% when we 
allowed a 75ms timeout, and with a 100ms timeout the reliability was 100\%. 
We also see from Figure \ref{fig:lrw_dur_conf} that the average duration 
of the LRW operations increases as the timeout is reduced. This may at 
first seem surprising, until we consider the duration distribution shown on 
Figure \ref{fig:nocont_dist}. This figure shows the percentage of 
LRW operations that ended within a given time interval, as seen on the x-axis.

There are a few important observations we can make when we cross reference 
Figure \ref{fig:nocont_dist} with Table \ref{tab:rel_nocont}. First, 
note that an approximate equal percentage of the LRW operations for each 
timeout value ended within the 20-40ms interval. Obviously the timeout 
had no effect on these, as one would expect. The next large grouping occurs 
at the 60-80ms time interval, where we find most of the operations 
from the 100ms timeout trial, and many from the 75ms trial. However, 
the 50ms trial is virtually absent from this interval, while it is instead 
almost the sole occupant in the 100-120ms interval. Since this 
interval includes the double of the timeout value, it is natural to 
hypothesize that for the majority of the LRW operations from the 
50ms trial that ended within this interval, $\textsf{T}_{Timeout}$ 
expired twice.  (To confirm this hypothesis, we investigated the behavior
of the cycle of broadcast (or rebroadcast triggered by $\textsf{T}_{Response}$ expiring) 
and acknowledgments, detailed in the next paragraph.)
A similar effect is seen for the 75ms trial at the 140-160ms interval. 
The only difference is that in this case we note that while approximately 
10\% of the LRW operations ended within this interval, the reliability data 
implies only 3\% of the operations failed. On closer inspection of the 
data we observed that most of the operations that ended within this interval 
were \textit{canceled}, not \textit{failed}. In either of the two trials, 
we see that reducing the timeout value had a negative effect not only on 
the reliability, but also on the duration of the LRW operations.


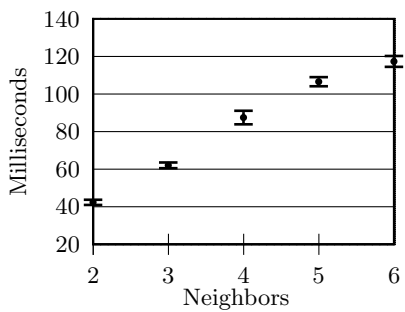
\begin{figure}[ht]
\begin{minipage}{\columnwidth}
\readdata{\Data}{avg_time_neigh_cont.csv}
\psset{dotscale=0.75}%
\psset{xunit=1cm,yunit=.025cm}
\psset{yticksize=0 4cm}
\begin{center}
\begin{pspicture}(1.8,-10)(6,140)
	{\footnotesize
	\psgrid[subgriddiv=0,griddots=10,gridlabels=0](2,20)(6,140)
  \psframe[fillstyle=solid,fillcolor=white,linestyle=solid](2,20)(6,140)
  \psaxes[Dy=20,Ox=2,Oy=20](2,20)(6,140)
  \rput{90}(1,80){Milliseconds}
  \rput(3.9,-9){Neighbors}
  \def\DoCoordinate#1#2{\psdot(#1,#2)}%
  \GetCoordinates{\Data}
  }
\end{pspicture}
\end{center}
\end{minipage}
\caption{Average LRW operation duration with 95\% confidence interval.}
\label{fig:ser_dur_conf} 
\end{figure}

\begin{figure}[ht]
\begin{minipage}{\textwidth}
\input{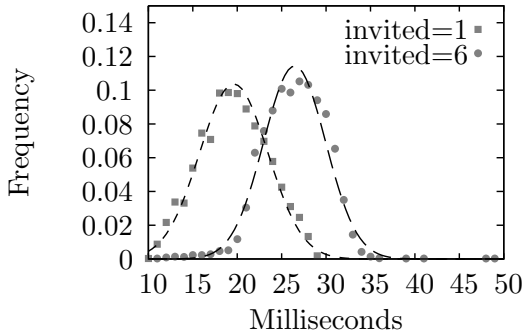}
\end{minipage}
\caption{Timings by Invitee List Size}
\label{fig:inviteeTimes}
\end{figure}

\begin{figure}[ht]
\begin{minipage}{\textwidth}
\input{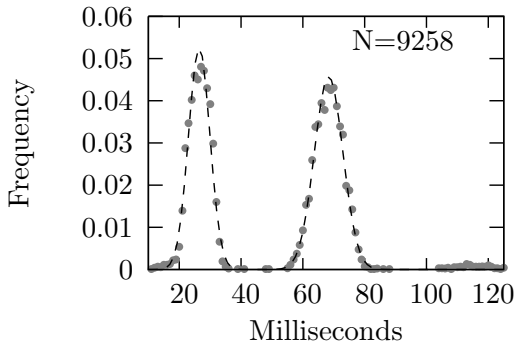}
\end{minipage}
\caption{Duration for $\textsf{T}_{Timeout}=\infty$}
\label{fig:totDur}
\end{figure}

In an experiment using seven TelosB motes, each mote acted as initiator
approximately 1,300 times, however the experiment tested only the first 
part of the protocol of Figure \ref{newfig-init}:  effectively 
$\textsf{T}_{Timeout}=\infty$ for this experiment.  
Each initMsg broadcast included the list of
neighbors who had not yet responded to the LRW operation.  The initiator's 
neighborhood was fixed to be the other six motes.  Thus, the initial broadcast of
an initMsg contained an invitee list of length six;  rebroadcasts contained smaller 
invitee lists.  An operation terminated at the instant the initiator had received
acknowledgments from all neighbors.  The experiment allocated approximately four
seconds to each LRW operation, to ensure that no contention between operations 
could occur.  In a total of 9,258 operations, 4,228 of them (45.6\%) were ``lucky'', 
that is, acknowledgments from all six neighbors occurred immediately following 
the initial broadcast, so that no rebroadcast was necessary.  The distribution
of operation times for these cases is shown in Figure \ref{fig:inviteeTimes},
indicated by $invited=6$.  Some 3,288 of the operations included a rebroadcast
containing a singleton invitee list, shown as $invited=1$ in the figure.  
The mean (standard deviation) operation duration in milliseconds for $invited=6$ 
was 26.5 (3.49), while for $invited=1$ the results are 19.6 (3.85).  The curves
in Figure \ref{fig:inviteeTimes} are Gaussian distributions fitted the the 
mean and standard deviations (we also obtained data for $invited=k$, $1<k<6$, 
which follow similar patterns).  We also measured the total duration (from 
start to termination) of each operation.  This is shown in Figure \ref{fig:totDur}, 
clearly reflecting $\texttt{T}_{Response}$ expiration and rebroadcast.  
The mean number of (re-) broadcasts per operation was 1.55, and the bimodal 
distribution in the figure explains this (actually, the distribution has three
modes, however the third does not contribute significantly).  These experiments
with $\textsf{T}_{Timeout}=\infty$ explain the results of 
Figure \ref{fig:nocont_dist} and Table \ref{tab:rel_nocont}.  One could
even use distributions suggested by the fitted curves to build an analytical
model, however this model would be impractical for situations where initiators
may contend with network traffic and 
for nonoptimistic cases where an LRW operation should be canceled.

\paragraph{}
Our second set of experiments introduced contention by having six initiators 
present in the graph. Recall that we enforced a topology on the network 
such that every initiator shares at least one potential neighbor with another 
initiator. In these experiments, each trial consisted of approximately 1000 series.

Due to the dynamic neighborhood, the number of motes attached to a 
single initiator varied over the course of one trial. Figure \ref{fig:ser_dur_conf} 
shows the average LRW operation duration with a 95\% confidence interval, 
depending on the size of the neighborhood. As we see the average duration 
has increased, especially for larger neighborhoods, compared to Figure \ref{fig:broad_uni_nocont}.

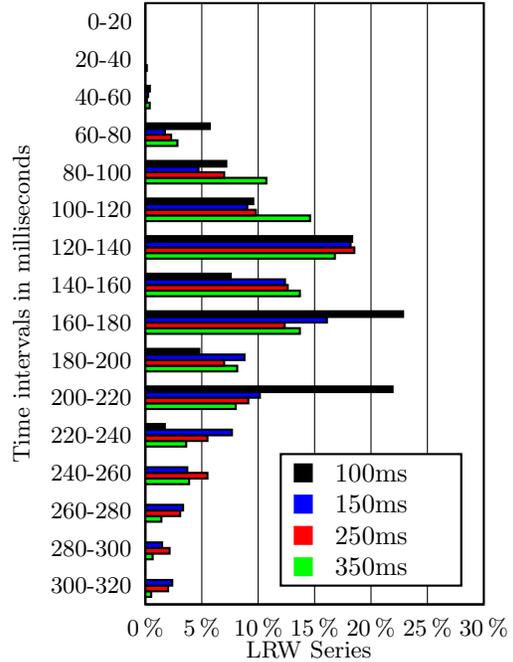
\begin{figure}[ht]
\begin{minipage}{\columnwidth}
\psset{xunit=0.15cm,yunit=0.5cm}
\psset{xticksize=0 8cm}
\def\pshlabel#1{#1\,\%}
\begin{center}
\begin{pspicture}(-2,-1)(30,16)
	{\footnotesize
	\psframe[fillstyle=solid,fillcolor=white,linestyle=solid](0,0)(30,16)
	\psaxes[ticks=x,labels=x,Ox=0,Oy=0,Dx=5,Dy=1](0,0)(30,16)
	\readpsbardata[header=true]{\data}{time_int_cont.csv}
	\psbarchart[barstyle={green,red,blue,black},orientation=horizontal]{\data}
  \rput(14.5,-1.2){LRW Series}
  \rput{90}(-11,8){Time intervals in milliseconds}
  
  \pspolygon[fillcolor=white,fillstyle=solid](12,4)(12,0.5)(28,0.5)(28,4)
  
  \psdots[linecolor=black,dotstyle=square*,dotsize=8pt](14,3.5)
	\rput(20,3.5){\color{black}{\small 100ms}}
	
	\psdots[linecolor=blue,dotstyle=square*,dotsize=8pt](14,2.667)
	\rput(20,2.667){\color{black}{\small 150ms}}
	
  \psdots[linecolor=red,dotstyle=square*,dotsize=8pt](14,1.833)
	\rput(20,1.833){\color{black}{\small 250ms}}
	
  \psdots[linecolor=green,dotstyle=square*,dotsize=8pt](14,1)
	\rput(20,1){\color{black}{\small 350ms}}
	}
\end{pspicture}
\end{center}
\end{minipage}
\caption{Distribution of series duration.}
\label{fig:time_interval_cont}
\end{figure}

\paragraph{}
Previously we examined the effect that reducing the timeout value would have on 
LRW duration and reliability in a non-contention network. We repeated 
these experiments with contention, starting with a timeout of 100ms and 
increasing it by 50ms in each trial. Table \ref{tab:rel_cont} shows the 
operation and series reliability of each trial, and 
Figure \ref{fig:cont_op_dur_dist} shows the distribution of the 
duration for LRW operations. Figure \ref{fig:cont_op_dur_dist} 
omits the data from the 200 and 300ms trials, as these were similar to the 250 and 350ms trials.

As seen in Table \ref{tab:rel_cont}, the operation reliability is very poor 
in the 100ms trial. But it increases rapidly, becoming approximately 
99 \% in the 200ms trial and 100 \% at 350ms. If we cross reference this 
with Figure \ref{fig:cont_op_dur_dist} we see that in the 100ms trial almost 
every LRW operation had completed within 220ms, while in the 350ms trial 
every transaction had completed within 320ms.

Due to the inclusion of contention, clock synchronization, and dynamic neighborhoods, 
the behavior of the protocol is significantly more complex than for previous trials. 
However we still notice the same basic trend that was evident in 
Figure \ref{fig:nocont_dist}: the percentage of operations that ended within 
160ms is largely similar for each trial, while at later intervals we notice 
first a drop, and then an increase in the number of operations from the 100ms trial. 

\begin{table}[ht]
\begin{minipage}{\columnwidth}
\begin{center}
{\footnotesize
\begin{tabular}{|r|c|c|}
\hline
 & Operation & Series\\
Timeout & Reliability & Reliability\\
\hline
100ms & 87.91\% & 55.11\% \\ 
\hline
150ms & 96.65 \% & 83.66 \% \\ 
\hline
200ms & 99.37 \% & 96.64 \% \\ 
\hline
250ms & 99.61 \% & 97.85 \% \\ 
\hline
300ms & 99.85 \% & 98.90 \% \\ 
\hline
350ms & 100.00 \% & 100.00 \% \\ 
\hline
\end{tabular}
}
\end{center}
\end{minipage}
\caption{Reliability with contention}
\label{tab:rel_cont} 
\end{table}

This becomes much more pronounced when we consider the series reliability 
in the third column in Table \ref{tab:rel_cont} and the duration distribution 
in Figure \ref{fig:time_interval_cont}. Here we clearly see that a 
large number of the series in the 100ms trial ended between 160 and 220ms.

\begin{figure}[ht]
\begin{minipage}{\columnwidth}

\readdata{\Data}{rel_cont.csv}
\psset{dotscale=0.75}%
\psset{xunit=0.015cm,yunit=.08cm}
\psset{yticksize=0 3.75cm}
\begin{center}
\begin{pspicture}(100,135)(350,185)
	{\footnotesize
  \psframe[fillstyle=solid,fillcolor=white,linestyle=solid](100,140)(350,180)
  \psaxes[Dy=5,Ox=100,Dx=50,Oy=140](100,140)(350,180)
  \rput{90}(40,160){Milliseconds}
  \rput(220,130){Timeout in milliseconds}
  \def\DoCoordinate#1#2{\psdot(#1,#2)}%
  \GetCoordinates{\Data}
  }
\end{pspicture}
\end{center}
\caption{Average series duration with 95\% confidence interval for varying timeout.}
\label{fig:rel_cont} 
\end{minipage}
\end{figure}
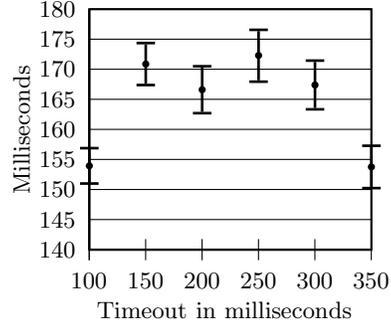

\paragraph{}
Recall that we observed from Figure \ref{fig:nocont_dist} that reducing 
the timeout value has the effect of increasing the duration of the 
LRW operations. Figure \ref{fig:rel_cont}, which shows the average 
series duration and 95\% confidence interval for each of the above trials, 
illustrates the same tendency. However, note that contrary to what we 
saw in Figure \ref{fig:nocont_dist}, when any reduction of the timeout 
value beyond 100ms resulted in an increased duration, we see in this 
figure that the series in the 100ms trial has approximately the same 
average duration as in the 350ms trial.

\section{Conclusions} \label{section:conclusion}

This paper's proposal and investigation of the LRW protocol  
is an attempt to answer the question:  what basic, single 
communication-round primitive maximizes the work accomplished
for a local neighborhood operation?  In some sense, an LRW
operation is a communication rendezvous, where members of 
a local neighborhood change states jointly.  Not surprisingly,
such an operation is more powerful than simpler primitives, 
such as neighborhood queries or unconditional broadcast-write
commands.  However, the advantages of LRW presume an 
optimistic, or speculative programming approach:  if 
contention is high or the semantics of the application do not
favor success, then LRW could be less attractive.  

Without notions of stable storage and journaling, which support ACID 
properties of database managers, consistency of LRW operations 
cannot be guaranteed;  however tuning the $\textsf{T}_{Timeout}$ 
parameter appropriately does increase the probability of non-failed operations.
The importance of consistency for abstractions like LRW, write-all, or
local transactions, is diminished for most wireless sensor network applications 
--- lowering component cost and conserving power have high priority.  Applications
can often be designed to tolerate some small probability of data inconsistency, 
using standard techniques of replication, filtering, outlier removal, and 
model-generated prediction.

Our experiments show that tuning $\textsf{T}_{Timeout}$ impacts 
both average duration and reliability.
Of the two most significant timers used, $\textsf{T}_{Timeout}$ 
and $\textsf{T}_{Response}$, we limited ourselves to studying the former; 
$\textsf{T}_{Response}$ was set to 40 milliseconds in all experiments.
A possible area of future research would be to examine the length 
of $\textsf{T}_{Response}$, and its effect on the protocol. There 
are many factors that would need to be considered in this case, such 
as size of neighborhood, size of the network, radio activity, and so on. 
Similar as for $\textsf{T}_{Timeout}$, we expect that there will be 
trade-offs between throughput and the number of unnecessary retransmissions.

One aspect of LRW we did not explore in this paper is the ``convenience'' of
LRW for common applications of sensor network programming.  The introduction
explains how LRW can be used for local data collection;  artificial examples
of consensus or leader election can easily be shown, however practical 
case studies would be helpful to evaluate LRW as a programming primitive.

\end{document}